\documentclass[conference]{IEEEtran}
\IEEEoverridecommandlockouts

\usepackage{cite}
\usepackage{algpseudocode}
\usepackage{graphicx}
\usepackage{amsmath}
\usepackage{amsfonts}
\usepackage{epstopdf}
\usepackage{xcolor}
\usepackage{enumerate}
\usepackage{algorithm}
\usepackage{multirow}
\usepackage{transparent}
\usepackage{subfigure}
\usepackage{float}

\def\BibTeX{{\rm B\kern-.05em{\sc i\kern-.025em b}\kern-.08em
		T\kern-.1667em\lower.7ex\hbox{E}\kern-.125emX}}

\newtheorem{corollary}{Corollary}
\newtheorem{theorem}{Theorem}

\newtheorem{lemma}{Lemma}

\newcommand{\R}{\textnormal{R}}
\let\ss= \scriptscriptstyle
\begin{document}

\title{Analysis of Receiver Covered by Heterogeneous Receptors in Molecular Communications}

\author{\IEEEauthorblockN{Xinyu Huang\IEEEauthorrefmark{1}, Yuting Fang\IEEEauthorrefmark{2}, Stuart T. Johnston\IEEEauthorrefmark{2}, Matthew Faria\IEEEauthorrefmark{2}, Nan Yang\IEEEauthorrefmark{1}, and Robert Schober\IEEEauthorrefmark{3}}

\IEEEauthorblockA{\IEEEauthorrefmark{1}School of Engineering, Australian National University, Canberra, ACT 2600, Australia}
\IEEEauthorblockA{\IEEEauthorrefmark{2}University of Melbourne, Parkville,
VIC 3010, Australia}
\IEEEauthorblockA{\IEEEauthorrefmark{3}Friedrich-Alexander University Erlangen-N\"{u}rnberg, 91058 Erlangen, Germany}}

\maketitle

\begin{abstract}
This paper analyzes the channel impulse response of an absorbing receiver (RX) covered by multiple non-overlapping heterogeneous receptors with different sizes and arbitrary locations in a molecular communication system. In this system, a point transmitter (TX) is assumed to be uniformly located on a virtual sphere at a fixed distance from the RX. Considering molecule degradation during the propagation from the TX to the RX, the expected molecule hitting rate at the RX over varying locations of the TX is analyzed as a function of the size and location of each receptor. Notably, this analytical result is applicable for different numbers, sizes, and locations of receptors, and its accuracy is demonstrated via particle-based simulations. Numerical results show that (i) the expected number of absorbed molecules at the RX increases with an increasing number of receptors, when the total area of receptors on the RX surface is fixed, and (ii) evenly distributed receptors lead to the largest expected number of absorbed molecules.
\end{abstract}


\section{Introduction}

Molecular communication (MC) has emerged as a promising technology to facilitate micro-scale or nano-scale communications \cite{farsad2016comprehensive}. In MC, information is encoded into molecules that are released from a transmitter (TX). After being released, the molecules propagate in a fluid medium until they arrive at a receiver (RX). The RX then detects and decodes the information encoded in the molecules. Therefore, an accurate modeling of practical RXs is of great significance for the design and development of MC systems.

In biology, living cells recognize specific molecules via receptor-ligand interactions, which are fundamental for a cell to communicate with its neighbours and the entire organism \cite{guryanov2016receptor}. Specifically, a signal is delivered to the cell by binding of ligands to their complementary receptors, which results in a cascade of chemical reactions. Due to the complexity of modeling the entire process of signal delivery, the RX models considered in the MC literature are relatively simple, compared to the actual process. For example, the most widely-adopted RX models in MC are the passive RX, fully absorbing RX, and reactive RX \cite{jamali2019channel}, where the passive RX ignores receptor-ligand interactions, while the fully absorbing RX and reactive RX ignore the receptor size and assume an infinite number of receptors covering the entire RX surface. To enhance the practicality of RX models, the authors of \cite{akdeniz2018molecular} considered a partially absorbing RX and some recent studies, e.g., \cite{akkaya2014effect,ahmadzadeh2016comprehensive,sun2020expected,lotter2021saturating}, assumed the uniform distribution of a finite number of receptors on the RX surface, where all receptors have the same size. Specifically, the authors of \cite{akkaya2014effect} assumed that molecules are absorbed once they hit the receptors, and the authors of \cite{ahmadzadeh2016comprehensive,sun2020expected} assumed that a reversible reaction occurs once molecules hit the receptors. Very recently, \cite{lotter2021saturating} investigated the receptor occupancy induced by the competition of molecules when binding to the receptors in synaptic MC.

While interesting, the previous studies may not be accurate in practical scenarios when the receptors on the RX surface have different sizes and arbitrary locations. For example, receptors tend to form clusters on the RX surface at specific locations \cite{duke2009equilibrium}. Since a given ligand can activate all receptors within a cluster, the cluster can be regarded as a single larger receptor. In \cite{lindsay2017first}, the authors investigated a RX covered by receptors with different sizes and arbitrary locations under steady-state conditions and only derived the capacitance of the RX, but did not investigate the time-varying number of molecules absorbed by the RX. In this paper, we model receptors as absorbing patches (APs) and assume that there are multiple non-overlapping APs on the RX surface, where the APs have different sizes and arbitrary but fixed locations. When molecules hit the APs, they are absorbed by the RX. We further assume a point TX uniformly located on a virtual sphere at a fixed distance from the RX. By taking into account that molecules may degrade when they propagate from the TX to the RX, we investigate the expected channel impulse response (CIR) between the TX and the RX averaged over the varying locations of the TX when only the distance between the TX and RX is fixed. Here, the expected CIR is defined as the expected molecule hitting rate at the RX \cite{jamali2019channel}. 

Our major contributions can be summarized as follows. We derive the expected molecule hitting rate, the expected fraction of absorbed molecules, and the expected asymptotic fraction of absorbed molecules as time approaches infinity at a RX with multiple APs, where all expressions are functions of the size and location of each AP. The desired expressions allow us to investigate the impact of different numbers, sizes, and locations of APs on molecular absorption. Furthermore, we compare three types of APs distributions, namely, APs evenly distributed over the RX surface (i.e., the APs are equally spaced), APs randomly distributed over the RX surface, and APs located within a region on the RX surface. Particle-based simulations (PBSs) are used to verify the accuracy of our expressions. Our numerical results reveal that the expected number of absorbed molecules increases with an increasing number of APs, when the total area of the RX surface covered by APs is fixed. We also show that evenly distributed APs lead to a larger number of absorbed molecules than the other two types of AP distributions considered.

\section{System Model}

\begin{figure}[!t]
\begin{center}
\includegraphics[height=1.5in,width=0.61\columnwidth]{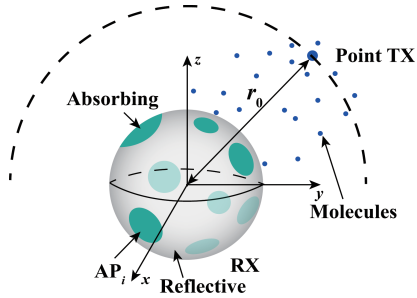}
\caption{Illustration of the MC system model where a point TX communicates with a spherical RX covered by multiple APs.}\label{sys}\vspace{-0.5em}
\end{center}\vspace{-4mm}
\end{figure}

In this paper, we consider an unbounded three-dimensional (3D) environment where a point TX communicates with a spherical RX, as depicted in Fig. \ref{sys}. We choose the center of the RX as the origin of the environment and denote the radius of the RX by $r_{\ss\mathrm{R}}$. There are $N_\mathrm{p}$ non-overlapping APs on the RX surface. We denote the $i$th AP by $\mathrm{AP}_i$. We assume the shape of each AP as a circle and denote $a_i$ as the radius of $\mathrm{AP}_i$. We define $\mathcal{A}$ as the ratio of the total area of APs to the RX surface, i.e., $\mathcal{A}=\sum_{i=1}^{N_\mathrm{p}}a_i^2/4r_{\ss\R}^2$. In spherical coordinates, we denote $\vec{l}_{i}=[r_{\ss\R}, \theta_{\mathrm{p},i}, \varphi_{\mathrm{p},i}]$ as the location of the center of $\mathrm{AP}_i$, where $\theta_{\mathrm{p},i}$ and $\varphi_{\mathrm{p},i}$ are the polar angle and azimuthal angle, respectively. Once molecules hit any AP, they are absorbed by the RX immediately. As is customary, for analytical tractability, we ignore the occupancy of molecules to the APs such that several molecules can be absorbed by an AP at the same time. Furthermore, we model the part of the RX surface that is not covered by APs as perfectly reflective, which means that molecules are reflected back once they hit this part of the RX surface.

In the considered system, the point TX is uniformly located on a virtual sphere with a distance $r_0$ from the center of the RX. For given $r_0$, the CIR of the RX is different for different locations of the TX. This is due to the fact that our considered RX has a heterogeneous boundary condition where the locations and sizes of the APs are determined by $\vec{l}_i$ and $a_i$, respectively. Accordingly, this work focuses on the expected CIR averaged over all possible locations of the TX when only the distance between the TX and RX is given. Thus, our results provide valuable insights for the practical case where only the distance between the TX and RX is known, while the accurate angular position of the TX relative to the RX is difficult to obtain. We note that the measurement of the distance between two cells is much easier than determining the relative angle between two cells by using the concentration gradient of molecules released by one of the cells \cite{lander2013cells}.

We assume that the propagation environment between TX and RX is a fluid medium with uniform temperature and viscosity. At time $t=0$, an impulse of $N_\sigma$ molecules of type $\sigma$ is released from the TX. Once molecules are released, they diffuse randomly with a constant diffusion coefficient $D_\sigma$. Moreover, we consider unimolecular degradation in the propagation environment, where type $\sigma$ molecules can degrade to type $\hat{\sigma}$ molecules that cannot be recognized by the RX, i.e., $\sigma\stackrel{k_\mathrm{d}}{\longrightarrow}\hat{\sigma}$ \cite[Ch. 9]{chang2005physical}, where $k_\mathrm{d}\;[\mathrm{s}^{-1}]$ is the degradation rate.

\section{Analysis of Channel Impulse Response}\label{acir}

In this section, we derive i) the expected molecule hitting rate, ii) the expected fraction of absorbed molecules, and iii) the expected asymptotic fraction of absorbed molecules as $t\rightarrow\infty$ at a RX with multiple APs by applying boundary homogenization \cite{dagdug2016boundary}. To this end, we first derive the expected CIR of a RX with a uniform surface reaction rate. A uniform surface reaction rate implies that the reactivity of the molecules is identical for all points on the RX surface. We then assume the system to be in steady state and derive the diffusion current of molecules across the RX surface. Based on the diffusion current in the steady state, we finally determine the effective reaction rate and apply it to derive the expected CIR of a RX with multiple APs.
\subsection{Problem Formulation}
In spherical coordinates, we denote the molecule concentration at time $t$ at location $\vec{r}$ by $C(\vec{r},t)$, $|\vec{r}|\geq r_{\ss\R}$. When an impulse of molecules is released from the TX at time $t=0$, the initial condition can be expressed as $C(\vec{r},t\rightarrow 0)=\delta(r-r_0)/(4\pi r_0^2)$ \cite[Eq. (3.61)]{schulten2000lectures},
where $\delta(\cdot)$ is the Dirac delta function. After the release, the diffusion of molecules in the propagation environment is described by Fick's second law as follows \cite{berg1993random}
\begin{align}\label{cr}
\frac{\partial C(\vec{r}, t)}{\partial t}=D_\sigma\nabla^2C(\vec{r}, t)-k_\mathrm{d}C(\vec{r}, t),
\end{align}
where $\nabla^2$ is the 3D spherical Laplacian. When molecules hit the RX surface, the reaction between the molecules and the RX surface is described by the radiation boundary condition \cite[Eq. (3.64)]{schulten2000lectures}, which is given by
\begin{align}\label{bc}
D_\sigma\frac{\partial C(|\vec{r}|, t)}{\partial |\vec{r}|}\bigg|_{|\vec{r}|=r_{\ss\R}}=wC(r_{\ss\R}, t),
\end{align}
where $w$ denotes the reaction rate. The unit of $w$ is $\mu\mathrm{m}/\mathrm{s}$, which is also validated by \eqref{bc}. We note that $w\rightarrow\infty$ when $\vec{r}\in\Omega_\mathrm{AP}$ while $w=0$ when $\vec{r}\in\Omega_\mathrm{R}$, where $\Omega_\mathrm{AP}$ and $\Omega_\mathrm{R}$ represent the parts of the RX surface that are fully covered and not covered by APs, respectively. Due to the heterogeneous boundary condition in \eqref{bc}, it is difficult to directly solve \eqref{cr} to obtain $C(\vec{r},t)$. In this paper, we apply boundary homogenization to derive the expected CIR. The main idea of boundary homogenization is replacing the heterogeneous boundary condition in \eqref{bc} by a uniform boundary condition with an appropriately chosen effective surface reaction rate, denoted by $w_\mathrm{e}$, which means that we replace the RX surface in Fig. \ref{sys} with an equivalent uniform surface with reaction rate $w_\mathrm{e}$. Hence, \eqref{bc} can be rewritten by replacing $w$ with $w_\mathrm{e}$. We derive the expected CIR of the RX and $w_\mathrm{e}$ in the following subsections.

\subsection{Expected CIR of RX with Uniform Surface Reaction Rate}

In this subsection, we analyze the expected CIR of a RX with uniform surface reaction rate $w$. Based on the initial condition in $C(\vec{r},t\rightarrow 0)$ and the boundary condition in \eqref{bc}, the authors of \cite{schulten2000lectures} derived $C(|\vec{r}|, t)$ by solving \eqref{cr} when $k_\mathrm{d}=0$. We denote $h_\mathrm{u}(t, w)$ as the expected molecule hitting rate at a RX with a uniform surface reaction rate. We specify $h_\mathrm{u}(t, w)$ including the effect of molecule degradation in the following lemma.
\begin{lemma}
The expected molecule hitting rate at a RX with uniform surface reaction rate $w$ at time $t$ is given by
\begin{align}\label{hu}
&h_\mathrm{u}(t,w)=\frac{r_{\ss\mathrm{R}}w}{r_0}\left[\frac{1}{\sqrt{\pi D_\sigma t}}\exp\left(-\frac{\varepsilon^2}{t}-k_\mathrm{d}t\right)-\gamma(w)\right.\notag\\
&\left.\times\exp\left[\gamma(w)(r_0-r_{\ss\mathrm{R}})+\zeta(w)t\right]
\mathrm{erfc}\left(\frac{\varepsilon}{\sqrt{t}}+\gamma(w)\sqrt{D_\sigma t}\right)\right],
\end{align}
where $\varepsilon\!=\!\frac{r_0-r_{\ss\mathrm{R}}}{\sqrt{4D_\sigma}}$, $\gamma(w)\!=\!\frac{wr_{\ss\mathrm{R}}+D_\sigma}{D_\sigma r_{\ss\mathrm{R}}}$,  $\zeta(w)\!=\!\gamma(w)^2D_\sigma-k_\mathrm{d}$, and $\mathrm{erfc}(\cdot)$ is the complementary error function.
\begin{IEEEproof}
According to \cite{heren2015effect}, $h_\mathrm{u}(t,w)$ can be obtained via $h_\mathrm{u}(t,w)=h_\mathrm{u}(t,w)\big|_{k_\mathrm{d}=0}\times\exp\left(-k_\mathrm{d}t\right)$. We also find that $h_\mathrm{u}(t,w)\big|_{k_\mathrm{d}=0}=4\pi r_{\ss\R}^2wC(r_{\ss\R}, t)\big|_{k_\mathrm{d}=0}$ based on \cite[Eq. (3.107)]{schulten2000lectures}. Combining these two results, we obtain \eqref{hu}.
\end{IEEEproof}
\end{lemma}

We further denote $H_\mathrm{u}(t,w)$ as the fraction of molecules captured by the RX by time $t$ and present it in the following corollary.
\begin{corollary}
The fraction of molecules captured by a RX with uniform surface reaction rate $w$ by time $t$ is given by
\begin{align}\label{Hu}
H_\mathrm{u}(t,w)=\frac{r_{\ss\mathrm{R}}w}{r_0}\left[\alpha_1(t)-\alpha_2(t,w)\right],
\end{align}
where
\begin{align}\label{d1}
\alpha_1(t)=&\frac{1}{2\sqrt{k_\mathrm{d}D_\sigma}}\left[\exp(-\beta)\mathrm{erfc}\left(\frac{\varepsilon}{\sqrt{t}}-\sqrt{k_\mathrm{d}t}\right)\right.\notag\\&\left.-\exp\left(\beta\right)\mathrm{erfc}\left(\frac{\varepsilon}{\sqrt{t}}+\sqrt{k_\mathrm{d}t}\right)\right]
\end{align}
and
\begin{align}\label{d2}		\alpha_2(t,w)\!=\!\frac{1}{2\zeta(w)}\left[\psi_1(t,w)-\psi_2(t,w)\right]-\frac{\gamma(w)\exp(-\beta)}{\zeta(w)}
\end{align}
with
\begin{align}\label{e1}		\psi_1(t,w)=&2\gamma(w)\exp\left(\gamma(w)(r_0-r_{\ss\mathrm{R}})+\zeta(w)t\right)\notag\\&\times\mathrm{erfc}\left(\frac{\varepsilon}{\sqrt{t}}+\gamma(w)\sqrt{D_\sigma t}\right),
\end{align}
\begin{align}\label{e2}		&\psi_2(t,w)=\left(\gamma(w)^2\sqrt{\frac{D_\sigma}{k_\mathrm{d}}}-\gamma(w)\right)\exp(-\beta)\notag\\
&\times\mathrm{erf}\left(\frac{\varepsilon}{\sqrt{t}}-\sqrt{k_\mathrm{d}t}\right)
-\left(\gamma(w)^2\sqrt{\frac{D_\sigma}{k_\mathrm{d}}}
+\gamma(w)\right)\notag\\
&\times\left[\exp(-\beta)-\exp(\beta)\mathrm{erfc}\left(\frac{\varepsilon}{\sqrt{t}}
+\sqrt{k_\mathrm{d}t}\right)\right],
\end{align}
$\mathrm{erf}(\cdot)=1-\mathrm{erfc}(\cdot)$ is the error function, and $\beta=(r_0-r_{\ss\mathrm{R}})\sqrt{k_\mathrm{d}/D_\sigma}$.
\end{corollary}
\begin{IEEEproof}
	$H_\mathrm{u}(t,w)$ can be obtained as $H_\mathrm{u}(t,w)=\int_{0}^{t}h_\mathrm{u}(u,w)\mathrm{d}u$. By substituting \eqref{hu} into this equation, we obtain \eqref{Hu}.
\end{IEEEproof}

We further denote $H_{\mathrm{u}, \infty}(w)$ as the asymptotic fraction of molecules captured by the RX as $t\rightarrow\infty$. We present $H_{\mathrm{u}, \infty}(w)$ in the following corollary.
\begin{corollary}\label{asy}
As $t\rightarrow\infty$, the asymptotic fraction of molecules captured by a RX with uniform surface reaction rate $w$ is given by
\begin{align}\label{Ha}
H_{\mathrm{u},\infty}(w)=\frac{r_{\ss\mathrm{R}}
w\left(\gamma(w)-\sqrt{\frac{k_\mathrm{d}}{D_\sigma}}\right)}{r_0\zeta(w)}\exp(-\beta).
\end{align}
\begin{IEEEproof}
Please see Appendix \ref{AA}.
\end{IEEEproof}
\end{corollary}

\subsection{Determination of Effective Reaction Rate}

In this section, we determine the effective reaction rate $w_\mathrm{e}$. First, we investigate the diffusion flux of molecules across the RX surface, denoted by $J$, in the steady state. The diffusion flux $[\textrm{molecule}\cdot\textrm{m}^{-2}\cdot\textrm{s}^{-1}]$ is the rate at which molecules move across a unit area in a unit time \cite{berg1993random}, and is given by \cite[Eq. (2.6)]{berg1993random}
\begin{align}
J=-D_\sigma\frac{\partial C(|\vec{r}|, t)}{\partial |\vec{r}|}\bigg|_{|\vec{r}|=r_{\ss\R}}.
\end{align}
We next define the rate of molecule movement across the RX surface in a unit time as the diffusion current, denoted by $I$, which is given by $I=-4\pi r_{\ss\mathrm{R}}^2J$. In the steady state, we have $\partial C(\vec{r},t)/\partial t=0$. If we set $k_\mathrm{d}=0$ in \eqref{cr}, we obtain
\begin{align}\label{nc}
\nabla^2C(\vec{r}, t)=0.
\end{align}

As explained in \cite{berg1977physics}, since \eqref{nc} is analogous to the Laplace's equation for the electrostatic potential in charge-free space, the diffusion current to an isolated absorbing RX of any size and shape can be expressed as \cite[Eq. (2.24)]{berg1993random}
\begin{align}\label{I}
I=4\pi D_\sigma GC_0,
\end{align}
where we define $G$ as the ``capacitance" of the RX and $C_0$ is the molecule concentration at $|\vec{r}|\rightarrow\infty$ in the steady state. We note that $C_0=1$ was adopted in some previous studies, e.g., \cite{lindsay2017first,ahmadzadeh2016comprehensive}. We further note that $G$ measures the ability of RX to absorb molecules, which is different from the conventional electrical capacitance of a conductor, denoted by $\hat{G}$. According to \cite{berg1993random}, if the RX and conductor have the same size and shape, $G$ can be obtained as $G=\hat{G}/4\pi\epsilon_0$, where $\epsilon_0$ is the vacuum permittivity. Since the expression for $\hat{G}$ has been derived for a variety of conductors, we can obtain $G$ directly if the RX and conductor have the same size and shape. For example, we have $\hat{G}=4\pi\epsilon_0r_{\ss\R}$ for a spherical conductor with radius $r_{\ss\R}$. According to the relationship between $G$ and $\hat{G}$, we can obtain the capacitance of a fully absorbing RX with the same radius, denoted by $G_\mathrm{a}$, as $G_\mathrm{a}=r_{\ss\mathrm{R}}$. Therefore, the diffusion current of a fully absorbing RX, denoted by $I_\mathrm{a}$, is given by $I_\mathrm{a}=4\pi D_\sigma r_{\ss\mathrm{R}}C_0$, which aligns with the derivation in \cite[Eq. (1)]{berg1977physics}.
For the RX with multiple APs in Fig. \ref{sys}, the capacitance of the RX, denoted by $G_\mathrm{p}$, was derived in \cite{lindsay2017first} by using the method of matched asymptotic expansions. Specifically, $G_\mathrm{p}$ is a function of the size and location of each AP, and it is given by \cite[Eq. (3.37a)]{lindsay2017first}
\begin{align}\label{Gc}
&\frac{1}{G_\mathrm{p}}=\frac{2}{N_\mathrm{p}\overline{m}\kappa r_{\ss\mathrm{R}}}\Bigg[1+\frac{\kappa}{2N_\mathrm{p}\overline{m}}
\ln\left(\frac{\kappa}{2}\right)\sum_{i=1}^{N_\mathrm{p}}m_i^2
+\frac{\kappa}{N_\mathrm{p}\overline{m}}\notag\\
&\times\bigg(\sum_{i=1}^{N_\mathrm{p}}m_is_i+2\sum_{i=1}^{N_\mathrm{p}}
\sum_{j=i+1}^{N_\mathrm{p}}m_im_j\mathcal{F}(\vec{l}_i',\vec{l}_j')\bigg)\!
+\!\left(\kappa\ln\left(\frac{\kappa}{2}\right)\!\!\right)^2\notag\\
&\times\frac{\vartheta}{4N_\mathrm{p}\overline{m}}
+\mathcal{O}\left(\kappa^2\ln\left(\frac{\kappa}{2}\right)\right)\Bigg],
\end{align}
where $\kappa=\frac{a_1}{r_{\ss\mathrm{R}}}$, $m_i=\frac{2a_i}{r_{\ss\mathrm{R}}\kappa\pi}$, $\overline{m}=\frac{1}{N_\mathrm{p}}\sum_{i=1}^{N_\mathrm{p}}m_i$, $s_i=\frac{m_i}{2}\left(\ln\left(\frac{4a_i}{r_{\ss\mathrm{R}}\kappa}\right)-\frac{3}{2}\right)$, $\vartheta=\frac{\left(\sum_{i=1}^{N_\mathrm{p}}m_i^2\right)^2}{N_\mathrm{p}\overline{m}}-\sum_{i=1}^{N_\mathrm{p}}m_i^3$, and
\begin{align}\label{F}
\mathcal{F}(\vec{l}_i', \vec{l}_j')=\left[\frac{1}{|\vec{l}_i'-\vec{l}_j'|}+\frac{1}{2}\ln|\vec{l}_i'-\vec{l}_j'|-\frac{1}{2}\ln\left(2+|\vec{l}_i'-\vec{l}_j'|\right)\right]
\end{align}
with $\vec{l}_i'=\vec{l}_i/r_{\ss\mathrm{R}}$. In \eqref{Gc}, $\mathcal{O}(\cdot)$ represents the infinitesimal of higher order, which is omitted during calculation.

When all APs have the same size, \eqref{Gc} can be further simplied as presented in the following corollary.
\begin{corollary}
When all APs have identical sizes, i.e., $a_1=a_2=\cdots=a_{\ss{N_\mathrm{p}}}$, $G_\mathrm{p}$ can be simplified as follows
\begin{align}\label{G1}
\frac{1}{G_\mathrm{p}}=&\frac{\pi}{N_\mathrm{p}\kappa r_{\ss\mathrm{R}}}\Bigg[1+\frac{\kappa}{\pi}\bigg(\!\ln(2\kappa)
-\frac{3}{2}+\frac{4}{N_\mathrm{p}}\sum_{i=1}^{N_\mathrm{p}}\sum_{j=i+1}^{N_\mathrm{p}}
\mathcal{F}(\vec{l}_i',\vec{l}_j')\bigg)\notag\\
&+\mathcal{O}\left(\kappa^2\ln\left(\frac{\kappa}{2}\right)\right)\Bigg].
\end{align}
\end{corollary}
\begin{IEEEproof}
	When $a_1=a_2=\cdots=a_{\ss{N_\mathrm{p}}}$, we have $m_i=\frac{2}{\pi}$ and $\vartheta=0$. By substituting $m_i$ and $\vartheta$ into \eqref{Gc}, we obtain \eqref{G1}.
\end{IEEEproof}

The capacitance of a RX with a single AP can be obtained by setting $N_\mathrm{p}=1$ in \eqref{G1}. In \cite{lindsay2017first}, the authors applied a high order asymptotic expansion to derive a more accurate expression, which is given by \cite[Eq. (6.31)]{lindsay2017first}
\begin{align}\label{g1a}
\frac{1}{G_\mathrm{p}}\bigg|_{N_\mathrm{p}=1}=&\frac{\pi}{\kappa r_\mathrm{R}}\bigg[1+\frac{\kappa}{\pi}\left(\ln(2\kappa)-\frac{3}{2}\right)-\frac{\kappa^2}{\pi^2}\left(\frac{\pi^2+21}{36}\right)\notag\\&+\mathcal{O}(\kappa^3\ln\kappa)\bigg].
\end{align}

The diffusion current of molecules across a RX with multiple APs, denoted by $I_\mathrm{p}$, can be obtained by replacing $G$ with $G_\mathrm{p}$ in \eqref{I}. We denote $h_\mathrm{p}(t)$, $H_\mathrm{p}(t)$, and $H_{\mathrm{p},\infty}$ as the expected molecule hitting rate, the expected fraction of absorbed molecules, and the expected asymptotic fraction of absorbed molecules at a RX with multiple APs, respectively. Exploiting the fact that the ratio between the expected asymptotic fraction of absorbed molecules at a RX with multiple APs and that at a fully absorbing RX is equal to the ratio between the diffusion current across a RX with multiple APs and that across a fully absorbing RX \cite{ahmadzadeh2016comprehensive}, we derive $w_\mathrm{e}$ and present $h_\mathrm{p}(t)$, $H_\mathrm{p}(t)$, and $H_{\mathrm{p},\infty}$ along with $w_\mathrm{e}$ in the following theorem.
\begin{theorem}
The expected molecule hitting rate, the expected fraction of absorbed molecules, and the expected asymptotic fraction of absorbed molecules as $t\rightarrow\infty$ at a RX with multiple APs are given by
\begin{align}\label{com}
h_\mathrm{p}(t)=h_\mathrm{u}(t,w_\mathrm{e}),~H_\mathrm{p}(t)=H_\mathrm{u}(t,w_\mathrm{e}),~ H_{\mathrm{p},\infty}=H_{\mathrm{u},\infty}(w_\mathrm{e}),
\end{align}
respectively, where
\begin{align}\label{wee}
w_\mathrm{e}=\frac{D_\sigma G_\mathrm{p}}{r_{\ss\mathrm{R}}(r_{\ss\mathrm{R}}-G_\mathrm{p})},
\end{align}
with $G_\mathrm{p}$ given in \eqref{Gc}. When all APs have the same size or there is only a single AP, $G_\mathrm{p}$ is given in \eqref{G1} and \eqref{g1a}, respectively.
\end{theorem}
\begin{IEEEproof}
	According to the relationship between the expected asymptotic fraction of absorbed molecules and the diffusion current, we have \begin{align}\label{ra}
		\frac{H_{\mathrm{u},\infty}(w_\mathrm{e})|_{k_\mathrm{d}=0}}
		{H_{\mathrm{a},\infty}}=\frac{I_\mathrm{p}}{I_\mathrm{a}},
	\end{align}
	where $H_{\mathrm{u},\infty}(w_\mathrm{e})\big|_{k_\mathrm{d}=0}
	=r_{\ss\mathrm{R}}^2w_\mathrm{e}/(r_0(w_\mathrm{e}r_{\ss\mathrm{R}}+D_\sigma))$ and $H_{\mathrm{a},\infty}$ is the asymptotic fraction of absorbed molecules as $t\rightarrow\infty$ at a fully absorbing RX, given by $H_{\mathrm{a},\infty}=r_{\ss\mathrm{R}}/r_0$ \cite[Eq. (3.116)]{schulten2000lectures}. By solving \eqref{ra}, we obtain \eqref{wee}. By substituting $w_\mathrm{e}$ into \eqref{hu}, \eqref{Hu}, and \eqref{Ha}, we obtain \eqref{com}.
\end{IEEEproof}

\section{Numerical Results}

\begin{figure}[!t]
\centering
\subfigure[Expected molecule hitting rate]{\begin{minipage}[t]{1\linewidth}
\centering
\includegraphics[width=2.6in]{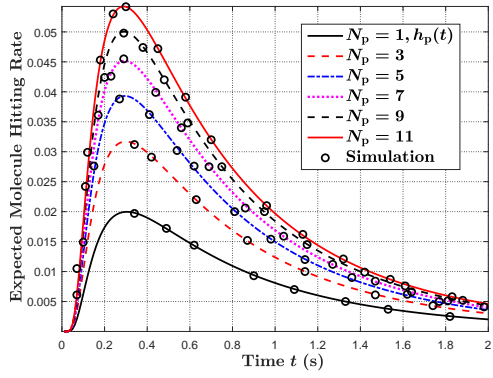}
\label{a}
\end{minipage}}
\quad
\subfigure[Expected number of absorbed molecules]{\begin{minipage}[t]{1\linewidth}
\centering
\includegraphics[width=2.6in]{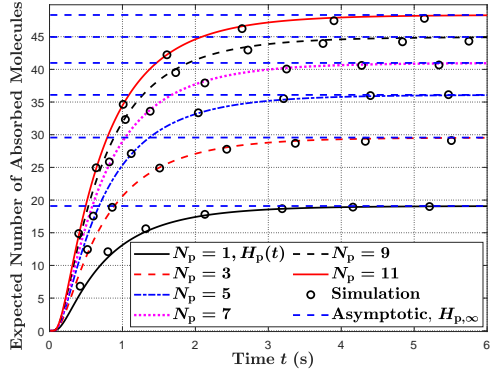}
\label{b}
\end{minipage}}
\centering
\caption{Expected molecule hitting rate at time $t$ and expected number of absorbed molecules until time $t$ versus time $t$ for varying $N_\mathrm{p}$, where $\mathcal{A}=0.05$.}\label{term}
\end{figure}

\begin{figure}[!t]
\centering
\subfigure[Expected molecule hitting rate]{\begin{minipage}[t]{1\linewidth}
\centering
\includegraphics[width=2.6in]{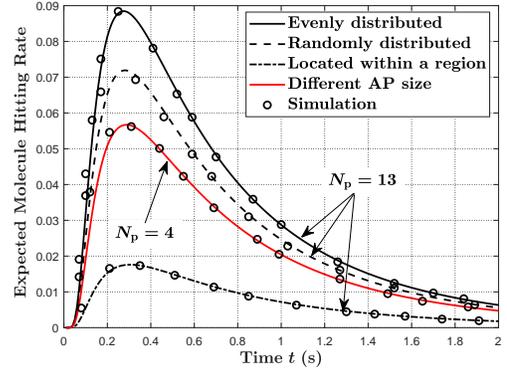}
\label{hr2}
\end{minipage}}
\quad
\subfigure[Expected number of absorbed molecules]{\begin{minipage}[t]{1\linewidth}
\centering
\includegraphics[width=2.6in]{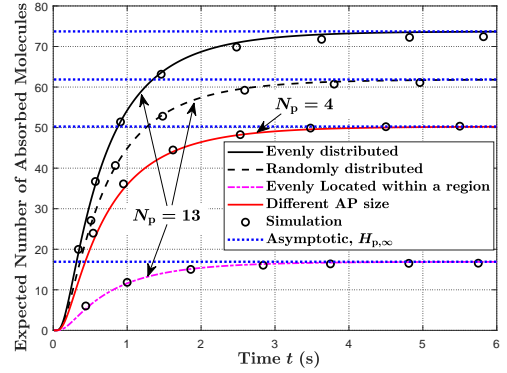}
\label{frac2}
\end{minipage}}
\centering
\caption{Expected molecule hitting rate at time $t$ and expected number of absorbed molecules at the RX until time $t$ versus time $t$ for varying locations and areas of APs, where $\mathcal{A}=0.1$.}\label{term2}
\end{figure}

In this section, we present numerical results to validate our theoretical analysis and offer insights for MC system design. Specifically, we use PBSs to simulate the random diffusion of the molecules, where we calculate the coordinates of the locations of molecules to be absorbed at the RX surface by using \cite[Eqs. (13)-(15)]{huang2021membrane}. If the locations of to-be-absorbed molecules are within the APs, we treat these molecule as molecules which have been absorbed. Otherwise, these molecules are reflected back to their positions at the start of the current simulation step \cite{ahmadzadeh2016comprehensive}. To model the location of the TX in the simulation, we fix the distance between the TX and the center of the RX as $r_0$ and change the location of the TX for each realization. The TX is located at any point on the virtual sphere with the same probability. The simulation time step is $\Delta t=10^{-6}\;\mathrm{s}$ and all results are averaged over 1000 realizations. Throughout this section, we set $\mathcal{A}=\{0.05, 0.1\}$ \cite{lindsay2017first}, $r_\mathrm{\ss\R}=10\;\mu\mathrm{m}$, $r_0=20\;\mu\mathrm{m}$, $D_\sigma=79.4\;\mu\mathrm{m}^2/\mathrm{s}$ \cite{yilmaz2014three}, $N_\sigma=1000$ \cite{heren2015effect}, $k_\mathrm{d}=0.8\;\mathrm{s}^{-1}$, and specify the value of $N_\mathrm{p}$ and the locations of the APs in each figure. In Fig. \ref{term} and Fig. \ref{term2}, we observe that the simulation results (denoted by markers) match well with the analytical curves (denoted by solid and dashed lines) generated based on Section \ref{acir}, which demonstrates the accuracy of our analysis.

In Fig. \ref{term}, we plot the expected molecule hitting rate at the RX at time $t$ in Fig. \ref{a} and the expected number of absorbed molecules at the RX by time $t$ in Fig. \ref{b}. We set $\mathcal{A}=0.05$ and increase the number of APs from 1 to 11. In this figure, all APs have the same size and are evenly distributed over the RX surface. Here, we apply the Fibonacci lattice \cite{gonzalez2010measurement} to determine the locations of the evenly distributed APs. Specifically, the location of $\mathrm{AP}_i$ is given by $\theta_{\mathrm{p},i}=\pi/2-\arcsin(2(i-\mathcal{B}-1)/N_\mathrm{p})$ and $\varphi_{\mathrm{p},i}=(4\pi(i-\mathcal{B}-1))/(1+\sqrt{5})$, where an integer $\mathcal{B}$ is given by $\mathcal{B}=(N_\mathrm{p}-1)/2$ and $N_\mathrm{p}$ is an odd number. In this figure, we observe that the expected hitting rate in Fig. \ref{a} and the expected number of absorbed molecules in Fig. \ref{b} increase with increasing $N_\mathrm{p}$. This is because APs can absorb molecules from any direction to the maximum when there are a larger number of APs on the RX surface, leading to an increased expected number of absorbed molecules.

In Fig. \ref{term2}, we vary the distributions and sizes of the APs on the RX surface. We set $\mathcal{A}=0.1$ and $N_\mathrm{p}=13$ for different AP distributions. Here, we consider three types of AP distributions, namely, APs evenly distributed on the RX surface, APs randomly distributed on the RX surface, and APs located within a region of the RX surface. For APs located within a region, we assume that APs are evenly distributed within $\theta_{\mathrm{p},i}\in[2.812, 3.471]$ and $\varphi_{\mathrm{p},i}\in[0, 2\pi]$. From this figure, we observe that evenly distributed APs achieve the highest expected hitting rate and the largest expected number of absorbed molecules, compared to the other two types of distributions. This is because evenly distributed APs ensure that the APs are equally spaced in all directions of the RX. Thus, for most locations of the TX, the probability of molecules hitting APs is the highest when the APs are evenly distributed, compared to the other two types of distributions. Moreover, we also consider $N_\mathrm{p}=4$ to validate our analytical expressions when the APs have different sizes. We denote $\mathcal{A}_{\mathrm{p}, i}$ as the ratio of the area of $\mathrm{AP}_i$ to the RX surface area and set $\mathcal{A}_{\mathrm{p}, 1}=0.01$, $\mathcal{A}_{\mathrm{p}, 2}=0.02$, $\mathcal{A}_{\mathrm{p}, 3}=0.03$, and $\mathcal{A}_{\mathrm{p}, 4}=0.04$. We also set the locations of the four APs to $\vec{l}_1=[10\;\mu\mathrm{m}, \pi/2, \pi]$, $\vec{l}_2=[10\;\mu\mathrm{m}, \pi/2, \pi/2]$, $\vec{l}_3=[10\;\mu\mathrm{m}, \pi/2, 0]$, and $\vec{l}_4=[10\;\mu\mathrm{m}, \pi/2, 3\pi/2]$.

\begin{figure}[!t]
	\begin{center}
		\includegraphics[height=2in,width=0.78\columnwidth]{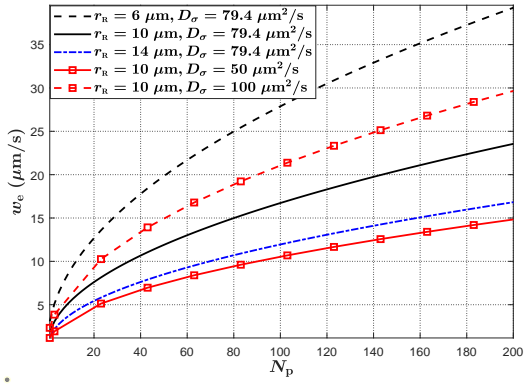}
		\caption{Effective reaction rate $w_\mathrm{e}$ versus $N_\mathrm{p}$ for varying $r_{\ss\R}$ and $D_\sigma$, where $\mathcal{A}=0.05$ and APs are evenly distributed over the RX surface.}\label{we}\vspace{-0.5em}
	\end{center}
	\vspace{-4mm}
\end{figure}

In Fig. \ref{we}, we plot $w_\mathrm{e}$ versus $N_\mathrm{p}$ for different values of $r_{\ss\R}$ and $D_\sigma$ to investigate the impact of $N_\mathrm{p}$, $r_{\ss\R}$, and $D_\sigma$ on the effective reaction rate $w_\mathrm{e}$. We note that a larger $w_\mathrm{e}$ means a higher probability of molecules hitting APs. We set $\mathcal{A}=0.05$ and consider evenly distributed APs on the RX surface. First, we observe that $w_\mathrm{e}$ increases when $N_\mathrm{p}$ increases, which indicates that the RX absorbs more molecules if there are more APs for a given $\mathcal{A}$. This observation aligns with the observations in Fig. \ref{term}. Second, we observe that $w_\mathrm{e}$ increases when $r_{\ss\R}$ decreases. This is because the AP-TX distances of different APs increase and become more similar when $r_{\ss\R}$ decreases and $r_0$ is kept fixed. Given that $\mathcal{A}$ is relatively small, the probability of molecules hitting APs becomes higher when the AP-TX distances increase and are more similar. Third, we observe that $w_\mathrm{e}$ increases when $D_\sigma$ increases. This is because the RX can absorb more molecules when the diffusion coefficient of molecules increases, which leads to higher $w_\mathrm{e}$.

\section{Conclusion}

In this paper, we proposed a RX model that is covered by multiple APs with different sizes and arbitrary locations. By considering a point TX uniformly located on a virtual sphere at a fixed distance from the RX, we derived closed-form expressions for the expected CIR between the TX and RX in the presence of molecule degradation. Simulations verified the accuracy of our analytical expressions. Our numerical results showed that when the proportion of the total area of APs to the RX surface is kept fixed, the expected number of absorbed molecules increases with the number of APs. They also showed that evenly distributed APs yield the largest expected number of absorbed molecules. Future directions for research include 1) investigating the optimal spatial and area arrangements of the APs for maximization of the expected number of absorbed molecules and 2) replacing the point TX with a practical TX model to characterize cell-to-cell communication.

\appendices

\section{Proof of Corollary \ref{asy}}\label{AA}

According to \eqref{Hu}, we have 
\begin{align}\label{H}
H_\mathrm{u,\infty}(w)=\frac{r_{\ss\mathrm{R}}w}{r_0}
\left[\alpha_1(\infty)-\alpha_2(\infty,w)\right].
\end{align}
Based on \eqref{d1}, we have $\alpha_1(\infty)=\exp(-\beta)/\sqrt{k_\mathrm{d}D_\sigma}$. To obtain $\alpha_2(\infty,w)$, we need to derive $\psi_1(\infty,w)$ and $\psi_2(\infty,w)$ in \eqref{d2}. In \eqref{e1}, if $\zeta(w)\leq0$, we have $\psi_1(\infty,w)=0$. If $\zeta(w)>0$, we obtain
\begin{align}\label{e1a}
&\psi_1\!(\!\infty,\!w\!)\!\!=\!\!2\gamma(w)\!\exp(\!\gamma(w)(r_0\!-\!r_{\ss\mathrm{R}}\!)\!)\!\!\lim_{t\rightarrow\infty}\!\!\frac{\mathrm{erfc}\!\left(\!\!\frac{\varepsilon}{\sqrt{t}}\!+\!\gamma(\!w\!)\sqrt{D_\sigma t}\!\right)}{\exp(-\zeta(w)t)}\notag\\&\overset{(a)}{=}\!\!2\gamma(w)\!\!\lim_{t\rightarrow\infty}\!\!\frac{\exp\!\!\left(\!-\!\!\left(\!\frac{\varepsilon^2}{ t}\!+\!k_\mathrm{d}t\!\right)\!\!\right)}{-\zeta(w)}\!\!\left(\!\!\frac{\varepsilon}{\sqrt{t^3}}\!-\!\gamma(w)\sqrt{\frac{D_\sigma}{t}}\right)\!\!=\!\!0,
\end{align}
where step $(a)$ is obtained by applying L'H{\^o}pital's rule. Therefore, we have $\psi_1(\infty,w)=0$ for any value of $\zeta(w)$. According to \eqref{e2}, we obtain $\psi_2(\infty,w)=-2\gamma(w)^2\sqrt{{D_\sigma}/{k_\mathrm{d}}}\exp\left(-\beta\right)$. By substituting $\psi_1(\infty,w)$ and $\psi_2(\infty,w)$ into \eqref{d2}, we obtain $\alpha_2(\infty,w)$. Then, by substituting $\alpha_1(\infty)$ and $\alpha_2(\infty,w)$ into \eqref{H}, we obtain \eqref{Ha}.

\bibliographystyle{IEEEtran}
\bibliography{ref2}

\end{document}